\documentclass[aps,prstab,superscriptaddress,longbibliography, twocolumn,amsmath,amssymb]{revtex4-2}
\usepackage{graphicx}
\usepackage{dcolumn}
\usepackage{hyperref}

\hypersetup{
    colorlinks=true,
    linkcolor=blue,
    citecolor=blue,
    filecolor=magenta,
    urlcolor=blue
}
\newcommand{\Figref}[1]{Fig.\ref{#1}}
\newcommand{\Tabref}[1]{Table~\ref{#1}}

\begin{document}

\title{Prism-based compensation of group delay dispersion in the components of a femtosecond laser resonator and analysis of the influence of prism configuration on laser radiation parameters}

\author{I.V. Beznosenko}
\altaffiliation[Corresponding author ]{\\Email address: beznosenko1989@gmail.com (I.V. Beznosenko)}

\author{A.V.~Vasyliev}
\author{G.V.~Sotnikov}
\author{A.I.~Povrozin}
\author{V.P.~Leshchenko}
\author{O.O.~Svystunov}

\address{National Science Center Kharkiv Institute of Physics and Technology \\
1, Akademichna St., Kharkiv, 61108, Ukraine}

\date{\today}

\begin{abstract}
A mathematical algorithm is described for calculating the minimum distance between prisms and the allowable prism insertion depth into the laser beam when solving the general problem of intracavity group delay dispersion (GDD) compensation. GDD is caused by the optical elements of the master oscillator in a femtosecond laser system. The calculation was carried out using analytical and geometrical methods. The influence of changing the distance between the prisms and the depth of their insertion into the laser beam on the GDD compensation has been investigated. Under experimental conditions, this changes the Q-factor of the laser resonator, which determines the central wavelength of the radiation and its spectral width in a titanium-sapphire laser.

\end{abstract}

\maketitle

\section{INTRODUCTION}
Femtosecond laser systems have a wide range of applications, including chemistry, metrology, materials science, telecommunication and data transmission, energy research, biology and medicine \cite{Sibbett2012}. They are also considered as promising sources of laser radiation for dielectric laser accelerators (DLA) \cite{Sotnikov2025,Vasiliev2023}.

\begin{figure}[!bh]
  \centering
  \includegraphics[width=0.96\textwidth]{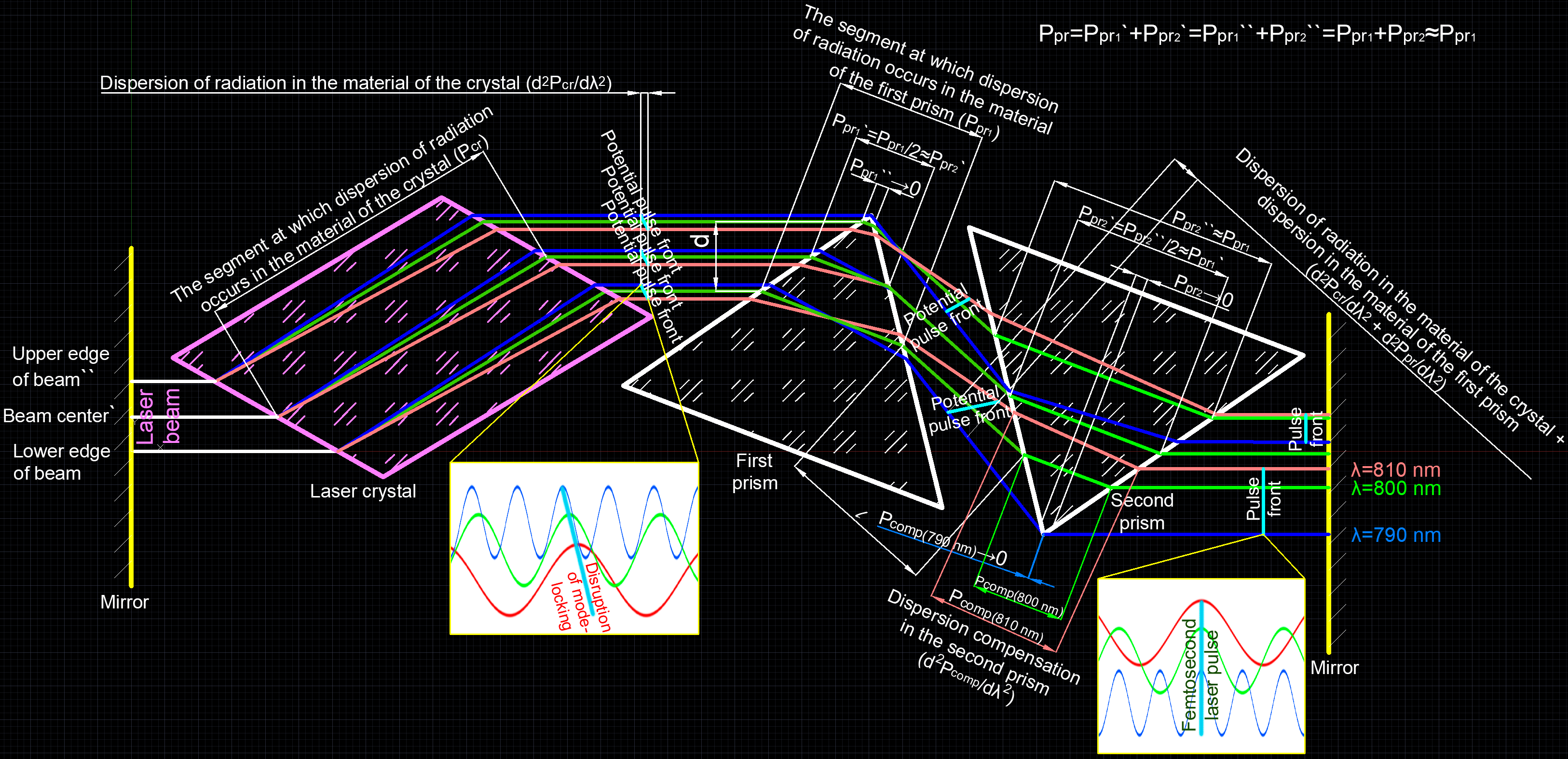}
  \caption{The principle of compensation of group delay dispersion (GDD) with minimal input of prisms into the laser beam: diameter of the laser beam (d), distance between prisms ($l$), path of the lower edge of the laser beam in the prism (P$_{pr}$), center of the beam (P$_{pr}'$), and the upper edge of the beam (P$_{pr}''$), path of the beam in the second prism, compensating GDD in the materials of the crystal and prisms for different spectral components of the laser radiation (P$_{comp}$), wavelength of individual spectral components ($\lambda$).}
  \label{Fig:01}
\end{figure}

The physical principle of femtosecond pulse generation is described in \cite{Spence1991}, while the mathematical theory of mode-locking, which is essential for generating high-power femtosecond pulses, is presented in \cite{Haus1975}. In the laser gain medium (such as a Ti:sapphire crystal), which has a relatively broad gain bandwidth, group delay dispersion (GDD) leads to spatial splitting and tilting of the femtosecond pulse front due to different refractive indices for spectral components of the radiation (see \Figref{Fig:01}).

\onecolumngrid

\twocolumngrid

To achieve a stable mode-locking regime, the positive GDD arising from the optical elements of the laser resonator must be compensated by negative GDD. This negative GDD can be introduced using a prism-based dispersion compensator consisting of two (\Figref{Fig:01}) or four prisms, as described in \cite{Fork1984}. A method for calculating such a prism compensator was proposed in \cite{Zaitsev2010}; however, it assumes the prisms operate in the minimum deflection mode, with the beam incident precisely at the Brewster angle and with minimal prism insertion into the laser beam.

In practice, during alignment of the master oscillator components in a femtosecond laser system \cite{Vasiliev2018}, deviations from the Brewster angle are often inevitable. In such cases, it is more practical to keep the distance between prisms fixed and instead vary the insertion depth of the prisms into the beam, which alters the positive GDD introduced by the prism material.

The goal of this work is to develop a general mathematical algorithm that allows:
\begin{itemize}
\item calculation of the minimum distance between prisms for minimal prism insertion;
\item determination of the necessary excessive insertion depth of the prisms for a fixed distance between prisms exceeding the minimum for a given beam diameter;
\item specification of arbitrary angles of incidence (not limited to the Brewster angle);
\item consideration of arbitrary apex angles of the prisms used in the compensator.
\end{itemize}

The algorithm is implemented using both analytical and geometrical methods.

The laser radiation has a continuous spectrum ranging from 790 nm to 810 nm. \Figref{Fig:01} shows the propagation paths of three spectral components with wavelengths of 790 nm, 800 nm, and 810 nm, representing the lower edge, the central wavelength, and the upper edge of the spectrum, respectively. For the beam center, the 790 nm and 810 nm components after the first prism are not shown. The angular divergence of the spectral components in the figure is increased, and the distances between optical elements are reduced for clarity.

\section{PROBLEM FORMULATION}
Before calculating the group delay dispersion (GDD) in the laser resonator elements and determining the conditions for its compensation using a two-prism compressor (including the minimum distance between the prisms and the dependence of the required additional prism insertion on the additional spacing of prisms), auxiliary computations are required.

To model GDD mathematically and geometrically and to determine the refractive index $n(\lambda)$ of laser radiation for different wavelengths $\lambda$ in the crystal and prism materials, the Sellmeier representation is used \cite{Marcuse1980}:

\begin{equation}\label{eq:01}
{n^2} - 1 = \sum\limits_{j = 1}^3 {\frac{{{\lambda ^2} \cdot {B_j}}}{{{\lambda ^2} - \lambda _j^2}}} ,
\end{equation}

where the Sellmeier coefficients $B_j$ and $\lambda_j$ for the laser crystal (Ti:Al$_2$O$_3$) were taken from \cite{Dodge1986}, and for the prism material (fused silica) -- from \cite{Tan1998}. Their values are shown in \Tabref{Tabl_1}.

\begin{table}
   \centering
   \caption{The coefficients for a three-term Sellmeier expansion}
   \begin{tabular}{lcc}
   \hline
   \hline
       \textbf{Medium} & \textbf{$B_j$} & \textbf{$\lambda_j$ ($\mu$m)} \\
   \hline
   \hline
          Laser crystal & $B_1$ = 1.43134930 & $\lambda_1$ = 0.0726631 \\
          (Ti:Al$_2$O$_3$) & $B_2$ = 0.65054713 & $\lambda_2$ = 0.1193242 \\
          & $B_3$ = 5.34140210 & $\lambda_3$ = 18.0282510 \\
   \hline
          Prisms & $B_1$ = 0.69616630 & $\lambda_1$ = 0.0684043 \\
          (fused silica) & $B_2$ = 0.40794260 & $\lambda_2$ = 0.1162414 \\
          & $B_3$ = 0.89747940 & $\lambda_3$ = 9.8961610 \\
   \hline
   \hline
   \end{tabular}
   \label{Tabl_1}
\end{table}

The first and second derivatives of the refractive index are obtained by differentiating Eq. (\ref{eq:01}):

\begin{equation}\label{eq:02}
\frac{{dn}}{{d\lambda }} =  - \frac{{{\lambda _0}}}{{{n_0}}} \cdot \sum\limits_{j = 1}^3 {\frac{{\lambda _j^2 \cdot {B_j}}}{{{{\left( {\lambda _0^2 - \lambda _j^2} \right)}^2}}}} ,
\end{equation}

\begin{equation}\label{eq:03}
\begin{array}{l}
\frac{{{d^2}n}}{{d{\lambda ^2}}} = \frac{1}{{{n_0}}} \cdot \sum\limits_{j = 1}^3 {\frac{{\lambda _j^2 \cdot \left( {3 \cdot \lambda _0^2 + \lambda _j^2} \right) \cdot {B_j}}}{{{{\left( {\lambda _0^2 - \lambda _j^2} \right)}^3}}} - } \\
 - \frac{1}{{{n_0}}} \cdot {\left( {\frac{{dn}}{{d\lambda }}} \right)^2},
\end{array}
\end{equation}

where $n_0$ is the refractive index for the central wavelength $\lambda_0$.

\Figref{Fig:02} illustrates geometrical parameters used to calculate GDD compensation in the laser crystal. The angular divergence of the spectral components in the figure is increased, and the distances between optical elements are reduced for clarity.

The incidence angle $\alpha$ of laser radiation polarized in the plane of the figure onto the crystal is chosen equal to the Brewster angle to minimize reflection losses:

\begin{equation}\label{eq:04}
\alpha = \arctan(n_{0CR}),
\end{equation}

where $n_{0CR}$ is the refractive index of Ti:Al$_2$O$_3$ crystal at the central wavelength of the laser radiation.

Both prisms are inserted into the beam of diameter $d$ until the transmitted power no longer increases to ensure minimum GDD.

The incidence angle $\Phi_1$ on the first surface of the first prism is also set to the Brewster angle:

\begin{equation}\label{eq:05}
\Phi_1 = \arctan(n_{0PR}),
\end{equation}

where $n_{0PR}$ is the refractive index of the prism material (fused silica) at the central wavelength of the laser radiation.

The adjacent faces of the prisms are aligned to be parallel. In the calculations, it is acceptable to use the actual apex angle $\varepsilon$ of each prism. However, in the ideal case, the apex angle $\varepsilon$ is determined from the condition that the output angle $\Phi_2$ of the radiation with the central wavelength is equal to the incidence angle $\Phi_1$, which minimizes reflection losses (this corresponds to the case where the central wavelength is incident at the Brewster angle). The prisms are manufactured so that their apex angles satisfy the following condition:

\begin{equation}\label{eq:06}
\varepsilon = 2\cdot\arcsin\left(\frac{\sin (\arctan(n_{0PR}))}{n_{0PR}}\right).
\end{equation}

The remaining angles are calculated for the central wavelength of the radiation using the following formulas.

Refraction angle $\psi_1$:

\begin{equation}\label{eq:07}
\psi_1 = \arcsin\left(\frac{\sin \Phi_1}{n_{0PR}}\right),
\end{equation}

Incidence angle $\psi_2$ on the second face:

\begin{equation}\label{eq:08}
\psi_2 = \varepsilon - \psi_1,
\end{equation}

Output angle $\Phi_2$:

\begin{equation}\label{eq:09}
\Phi_2 = \arcsin(n_{0PR} \cdot \sin \psi_2).
\end{equation}

Its derivatives were found by differentiation \cite{Fork1984}:

\begin{equation}\label{eq:10}
\frac{{d{\Phi _2}}}{{d{n_{PR}}}} = \frac{{\sin {\psi _2} + \cos {\psi _2} \cdot \tan {\psi _1}}}{{\cos {\Phi _2}}},
\end{equation}

\begin{equation}\label{eq:11}
\frac{{{d^2}{\Phi _2}}}{{dn_{_{PR}}^2}} = \tan {\Phi _2} \cdot {\left( {\frac{{d{\Phi _2}}}{{d{n_{PR}}}}} \right)^2} - \frac{{{{\tan}^2}{\psi _1}}}{{{n_{0PR}}}} \cdot \frac{{d{\Phi _2}}}{{d{n_{PR}}}}.
\end{equation}

The angular divergence $\Delta\Phi_2$ of the radiation after passing through the first prism arises due to wave dispersion, which results from differences in the refractive index for various spectral components of the laser pulse. It is calculated using the following expression \cite{Zaitsev2010}:

\begin{equation}\label{eq:12}
\Delta {\Phi _2} = 2 \cdot \Delta \lambda  \cdot \frac{{d{n_{PR}}}}{{d\lambda }},
\end{equation}

where $\Delta\lambda = \lambda_{max} - \lambda_{min}$ is the spectral width of the laser radiation. In our study, $\Delta\lambda$ = 810 nm - 790 nm = 20 nm. $\frac{dn_{PR}}{d\lambda}$ is calculated using Eq. (\ref{eq:02}), substituting $n_{0PR}$ obtained from Eq. (\ref{eq:01}) in place of $n_0$, and using the $B_j$ and $\lambda_j$ coefficients from \Tabref{Tabl_1} for fused silica. Similarly, the value of $\frac{d^2n_{PR}}{d\lambda^2}$ is calculated from Eq. (\ref{eq:03}). \\

\section{GDD IN THE LASER CRYSTAL AND ITS COMPENSATION BY A PRISM PAIR}
Group delay dispersion (GDD) is directly related to the second derivative of the optical path of radiation in a material \cite{Fork1984}. Since a positive GDD arises in the laser crystal (denoted as P$_{CR}$ in \Figref{Fig:01}), it must be compensated by introducing a negative GDD using a prism pair. The main contribution to the negative GDD is made by the second prism.

However, taking into account the finite width of the laser beam, even with minimal insertion of the prisms into the beam, an additional positive GDD arises in the material of the first prism (segment P$_{PR1}$ in \Figref{Fig:01}). To compensate for this, the distance between the prisms must be increased.

With minimal prism insertion (considering the lower edge of the beam), the second prism introduces purely negative GDD. This is because the shortest wavelength spectral component of the radiation has a minimal optical path (P$_{COMP(790nm)}$ in \Figref{Fig:01}) in the second prism, which maximizes the path difference between the long- and short-wavelength components.

\begin{figure}[!bh]
  \centering
  \includegraphics[width=0.78\textwidth]{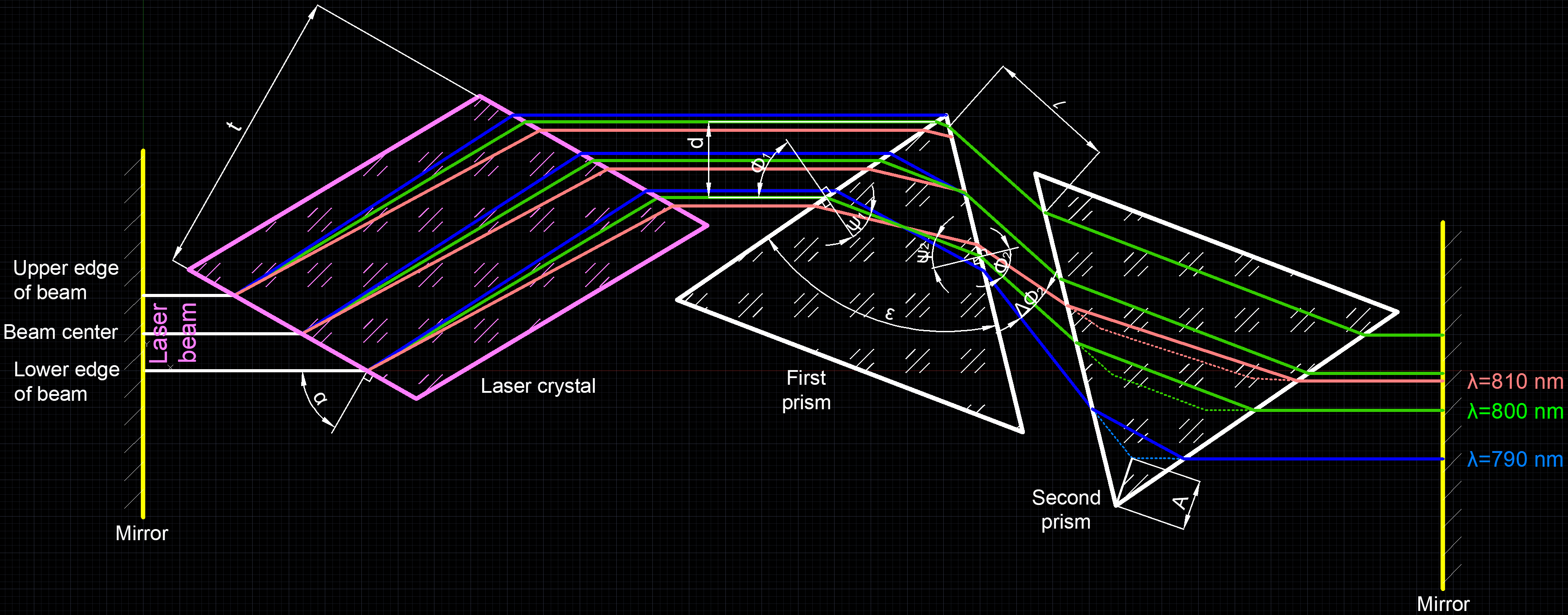}
  \caption{Geometrical parameters used to calculate GDD compensation using a two-prism system. Dashed lines indicate ray paths prior to additional insertion of the second prism by depth $A$.}
  \label{Fig:02}
\end{figure}

With excessive insertion of the second prism (depth $A$ in \Figref{Fig:02}), along with negative GDD, it also begins to introduce positive GDD similar to the first prism. This happens because the optical path of the short-wavelength component is no longer approaching zero.

To fully compensate GDD in the system, the following condition must be satisfied:

\begin{equation}\label{eq:13}
\frac{d^2P_{CR}}{d\lambda^2} + \frac{d^2P_{PR}}{d\lambda^2} + \frac{d^2P_{ADD\_PR}}{d\lambda^2} + \frac{d^2P_{COMP}}{d\lambda^2} = 0,
\end{equation}

where $\frac{d^2P_{CR}}{d\lambda^2}$ is the second derivative of the optical path of radiation in the laser crystal, $\frac{d^2P_{PR}}{d\lambda^2}$ is the second derivative of the optical path of radiation in the prisms at their minimal insertion into the laser beam, $\frac{d^2P_{ADD\_PR}}{d\lambda^2}$ is the second derivative of the additional optical path in the prisms caused by their excessive insertion into the laser beam, $\frac{d^2P_{COMP}}{d\lambda^2}$ is the second derivative of the optical path in the second prism for radiation that has undergone spectral spatial separation after passing through the first prism due to wave dispersion \cite{Fork1984,Zaitsev2010}:

\onecolumngrid

\begin{equation}\label{eq:14}
\frac{{{d^2}{P_{CR}}}}{{d{\lambda ^2}}} = \frac{t}{{{{\left( {n_{_{0CR}}^2 - {{\sin }^2}\alpha } \right)}^{3/2}}}} \cdot \left[ {{n_{0CR}} \cdot \left( {n_{_{0CR}}^2 - 2 \cdot {{\sin }^2}\alpha } \right) \cdot \frac{{{d^2}{n_{CR}}}}{{d{\lambda ^2}}} + \frac{{{{\sin }^2}\alpha  \cdot \left( {n_{_{0CR}}^2 + 2 \cdot {{\sin }^2}\alpha } \right)}}{{n_{_{0CR}}^2 - {{\sin }^2}\alpha }} \cdot {{\left( {\frac{{d{n_{CR}}}}{{d\lambda }}} \right)}^2}} \right],
\end{equation}

\begin{equation}\label{eq:15}
\frac{{{d^2}{P_{PR}}}}{{d{\lambda ^2}}} = \frac{{d \cdot \sin \varepsilon }}{{\cos {\Phi _1}}} \cdot \left[ \begin{array}{l}
\frac{1}{{\cos (\varepsilon  - {\psi _1})}} \cdot \left( {1 + \frac{{{\tan(\varepsilon  - {\psi _1})} \cdot \sin {\Phi _1}}}{{\sqrt {n_{_{0PR}}^2 - {{\sin }^2}{\Phi _1}} }}} \right) \cdot \frac{{{d^2}{n_{PR}}}}{{d{\lambda ^2}}} + \frac{{{{\sin }^2}{\Phi _1}}}{{{n_{0PR}} \cdot \cos (\varepsilon  - {\psi _1}) \cdot (n_{_{0PR}}^2 - {{\sin }^2}{\Phi _1})}} \times \\
 \times \left( {\frac{1}{{{{\cos }^2}(\varepsilon  - {\psi _1})}} + {{\tan}^2}(\varepsilon  - {\psi _1}) - \frac{{\tan(\varepsilon  - {\psi _1}) \cdot \sin {\Phi _1}}}{{\sqrt {n_{_{0PR}}^2 - {{\sin }^2}{\Phi _1}} }}} \right) \cdot {\left( {\frac{{d{n_{PR}}}}{{d\lambda }}} \right)^2}
\end{array} \right],
\end{equation}

\begin{equation}\label{eq:16}
\frac{{{d^2}{P_{ADD\_PR}}}}{{d{\lambda ^2}}} = \frac{{{\rm A} \cdot \cos \left( {\varepsilon /2 - {\Phi _1}} \right)}}{d} \cdot \frac{{{d^2}{P_{PR}}}}{{d{\lambda ^2}}},
\end{equation}

\begin{equation}\label{eq:17}
\frac{{{d^2}{P_{COMP}}}}{{d{\lambda ^2}}} =  - l \cdot \left\{ \begin{array}{l}
\sin \Delta {\Phi _2} \cdot \left[ {\frac{{{d^2}{n_{PR}}}}{{d{\lambda ^2}}}\left( { - \frac{{d{\Phi _2}}}{{d{n_{PR}}}}} \right) + {{\left( {\frac{{d{n_{PR}}}}{{d\lambda }}} \right)}^2} \cdot \left( { - \frac{{{d^2}{\Phi _2}}}{{dn_{_{PR}}^2}}} \right)} \right] + \\
 + \cos \Delta {\Phi _2} \cdot {\left( {\frac{{d{n_{PR}}}}{{d\lambda }}} \right)^2} \cdot {\left( { - \frac{{d{\Phi _2}}}{{d{n_{PR}}}}} \right)^2}
\end{array} \right\},
\end{equation}

\twocolumngrid

where $t$ is the width of the crystal, $d$ is the diameter of the laser beam when fully inserted into the prisms, $A$ is the total depth of excessive insertion of the prisms into the laser beam (\Figref{Fig:02}), $\frac{dn_{CR}}{d\lambda}$ is calculated using Eq. (\ref{eq:02}), substituting $n_{0CR}$ obtained from Eq. (\ref{eq:01}) in place of $n_0$, and using values of $B_j$ and $\lambda_j$ for Ti:Al$_2$O$_3$ from \Tabref{Tabl_1}. Similarly, $\frac{d^2n_{CR}}{d\lambda^2}$ is calculated from Eq. (\ref{eq:03}).

By substituting Eq. (\ref{eq:14})--(\ref{eq:17}) into (\ref{eq:13}), one can determine the distance $l$ between prisms (\Figref{Fig:02}) required for full compensation of GDD arising both in the crystal and in the prisms:

\onecolumngrid

\begin{equation}\label{eq:18}
l = \frac{{\frac{{{d^2}{P_{CR}}}}{{d{\lambda ^2}}} + \frac{{{d^2}{P_{PR}}}}{{d{\lambda ^2}}} + \frac{{{d^2}{P_{ADD\_PR}}}}{{d{\lambda ^2}}}}}{{\sin \Delta {\Phi _2} \cdot \left[ {\frac{{{d^2}{n_{PR}}}}{{d{\lambda ^2}}}\left( { - \frac{{d{\Phi _2}}}{{d{n_{PR}}}}} \right) + {{\left( {\frac{{d{n_{PR}}}}{{d\lambda }}} \right)}^2} \cdot \left( { - \frac{{{d^2}{\Phi _2}}}{{dn_{_{PR}}^2}}} \right)} \right] + \cos \Delta {\Phi _2} \cdot {{\left( {\frac{{d{n_{PR}}}}{{d\lambda }}} \right)}^2} \cdot {{\left( { - \frac{{d{\Phi _2}}}{{d{n_{PR}}}}} \right)}^2}}}.
\end{equation}

\twocolumngrid

Excessive insertion of the prisms effectively increases the laser beam diameter, which is accounted for in Eq. (\ref{eq:15}). This conditional beam expansion is described by Eq. (\ref{eq:16}) as $A \cdot \cos (\varepsilon/2 - \Phi_1)$ and is determined using the sine theorem. By substituting Eq. (\ref{eq:14})--(\ref{eq:17}) into (\ref{eq:13}), one can also find the total depth of excessive prism insertion at a known distance between prisms:

\begin{equation}\label{eq:19}
{\rm A} = \frac{{ - d \cdot \left( {\frac{{{d^2}{P_{CR}}}}{{d{\lambda ^2}}} + \frac{{{d^2}{P_{PR}}}}{{d{\lambda ^2}}} + \frac{{{d^2}{P_{COMP}}}}{{d{\lambda ^2}}}} \right)}}{{\cos \left( {\frac{\varepsilon }{2} - {\Phi _1}} \right) \cdot \frac{{{d^2}{P_{PR}}}}{{d{\lambda ^2}}}}}.
\end{equation}

\begin{figure}[!bh]
  \centering
  \includegraphics[width=0.48\textwidth]{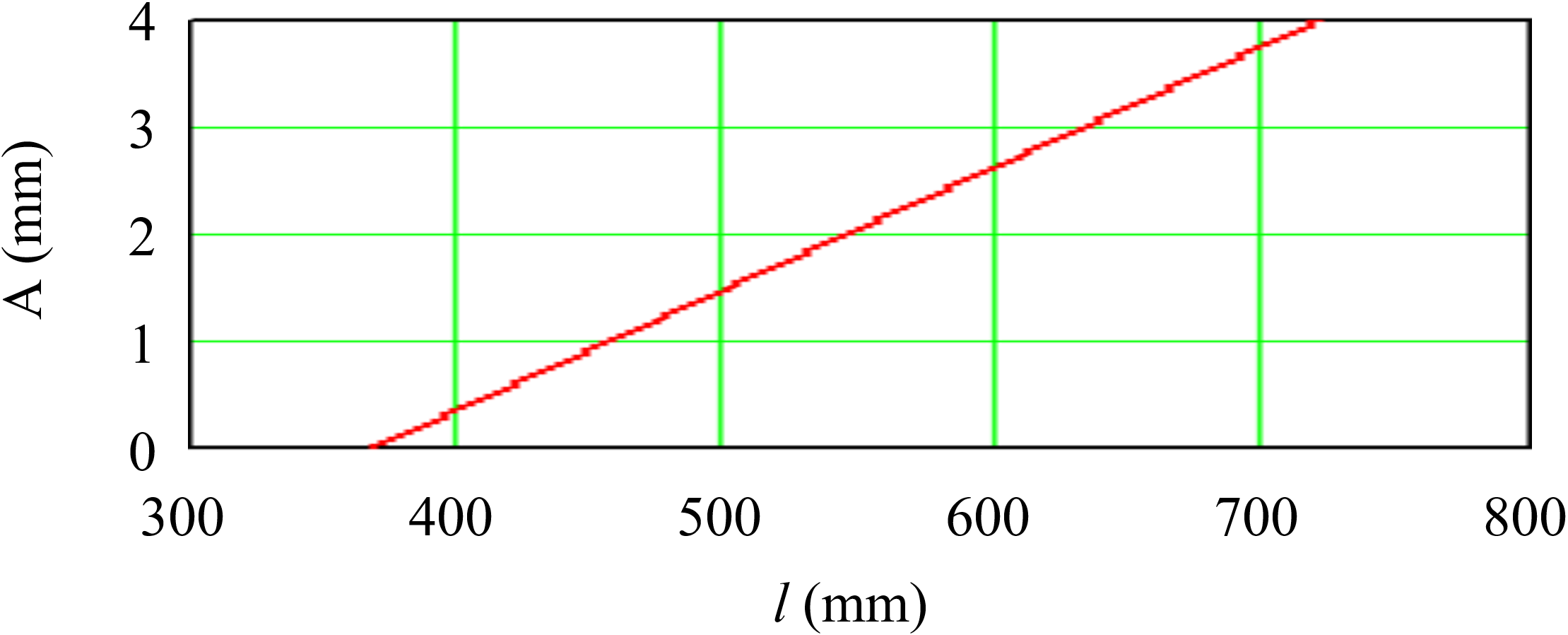}
  \caption{Dependence of excessive prism insertion $A$ on the distance $l$ between prisms.}
  \label{Fig:03}
\end{figure}

From Eq. (\ref{eq:19}), one can construct the dependence of the excessive prism insertion depth $A$ on the distance $l$ between prisms (\Figref{Fig:03}) for the case $\lambda_0$ = 800 nm, $\Delta\lambda$ = 20 nm, $t$ = 4.5 mm, $d$ = 2 mm.

For the given parameters, the minimum distance between prisms ensuring full GDD compensation without excessive insertion is 369 mm (see \Figref{Fig:03}). At shorter distances, the prisms must be partially withdrawn from the beam (reducing $d$), resulting in power loss. This means the value of $l$ at $A$ = 0 represents the minimum permissible distance. When $l$ is more than 369 mm, excessive insertion of the prisms into the beam is required.

\section{THE INFLUENCE OF PRISM CONFIGURATION ON LASER RADIATION PARAMETERS}
In our physical experiments, the distance between the prisms is fixed at approximately 610 mm, and changing this distance proved to be quite labor-intensive. At the same time, the depth of excessive insertion of the second prism into the laser beam could be easily adjusted within the range of 0 to 4 mm. Therefore, preliminary calculations were performed for the case of a fixed distance between prisms and variable depth of second prism insertion.

\begin{figure}[!bh]
  \centering
  \includegraphics[width=0.48\textwidth]{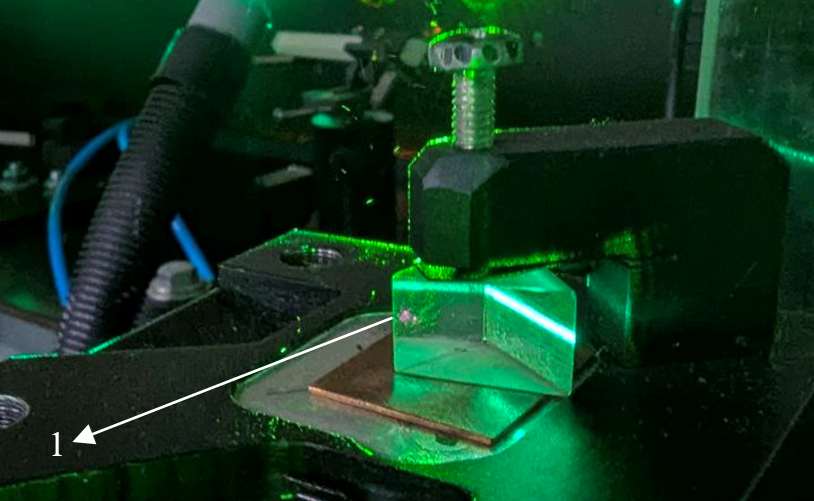}
  \caption{Photograph of a prism inserted into a laser beam (1).}
  \label{Fig:04}
\end{figure}

Photograph of a prism inserted into a laser beam is shown in \Figref{Fig:04}.

\begin{figure}[!bh]
  \centering
  \includegraphics[width=0.48\textwidth]{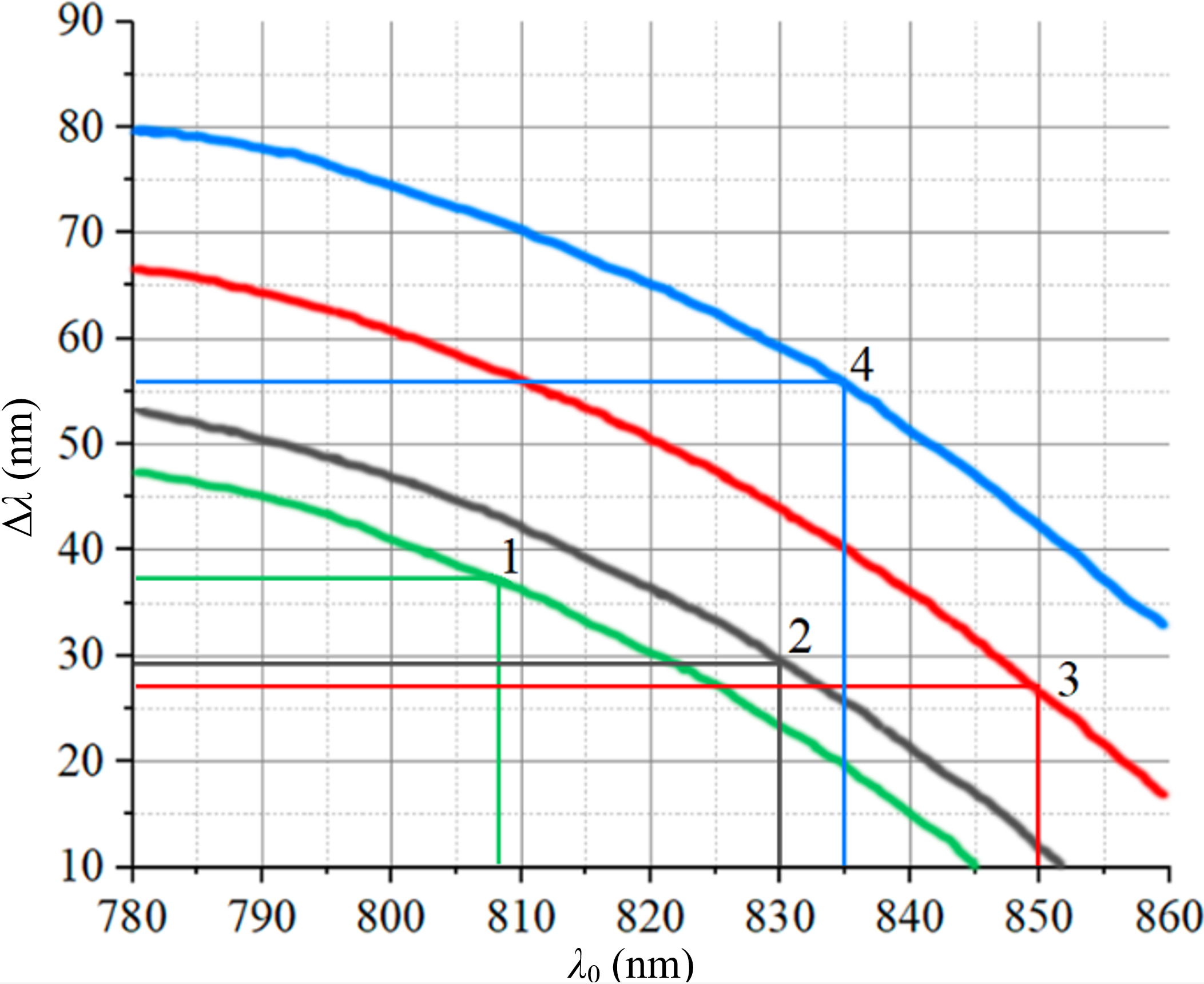}
  \caption{Calculated possible combinations of $\Delta\lambda$ and $\lambda_0$ in the mode-locking regime for $l$ = 610 mm and different values of $A$: 3.07 mm (1), 3.16 mm (2), 3.38 mm (3), 3.6 mm (4). Numbers 1--4 correspond to the experimental data in \Figref{Fig:06}.}
  \label{Fig:05}
\end{figure}

Our femtosecond laser system is capable of generating pulses with sufficient power for the planned dielectric laser acceleration (DLA) experiments within a central wavelength range of 780 nm to 860 nm and a spectral bandwidth range of 10 nm to 90 nm. As follows from Eq. (\ref{eq:13}), at certain values of $A$ and $l$, by reconfiguring the resonator elements, it is possible to achieve mode-locking and obtain radiation spectra with different combinations of central wavelength $\lambda_0$ and spectral bandwidth $\Delta\lambda$ on the curves shown in \Figref{Fig:05}.

\begin{figure}[!bh]
  \centering
  \includegraphics[width=0.48\textwidth]{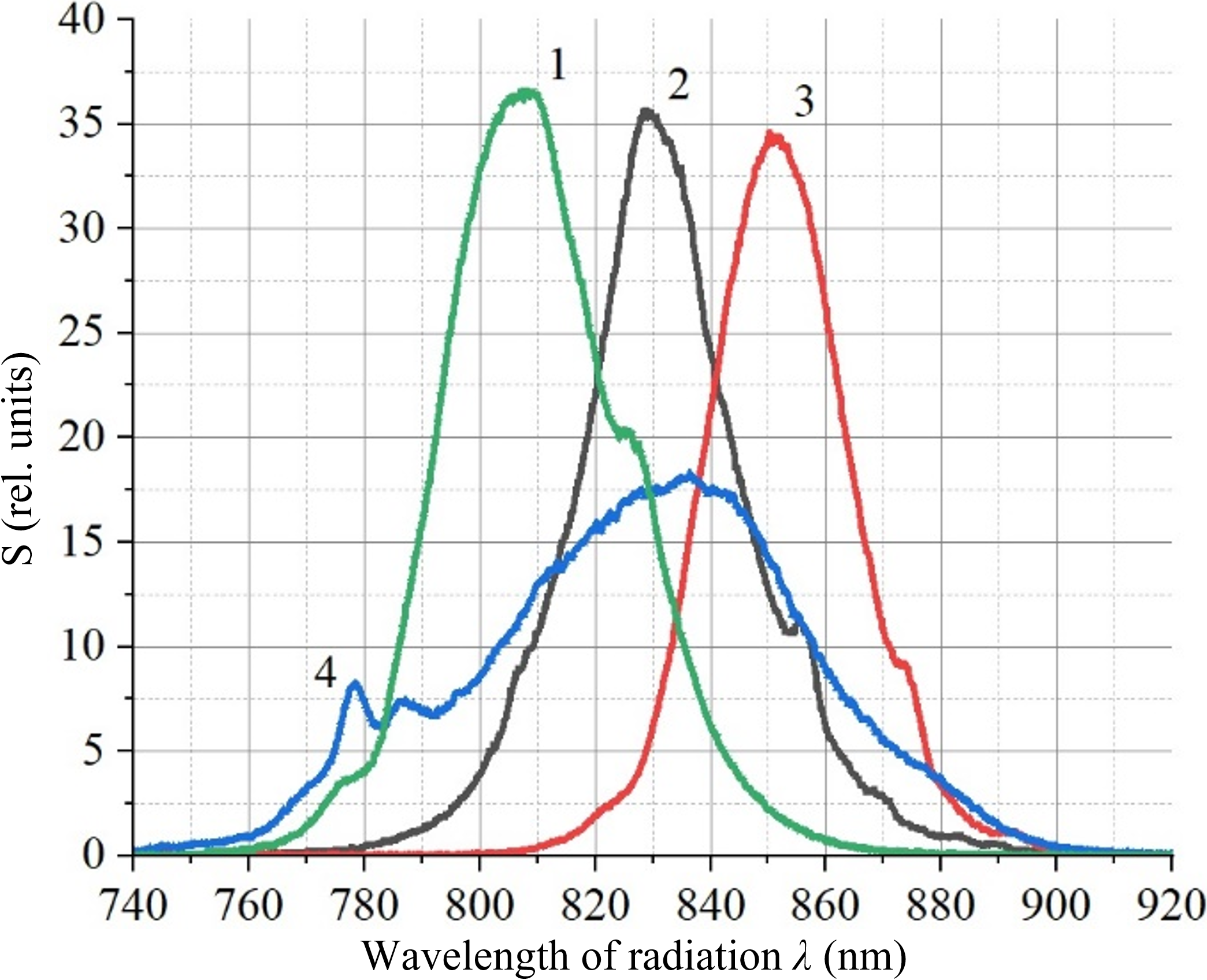}
  \caption{Experimental radiation spectrum in the mode-locking regime for different values of $A$: 3.07 mm (1), 3.16 mm (2), 3.38 mm (3), 3.6 mm (4).}
  \label{Fig:06}
\end{figure}

With a distance between prisms of 610 mm, we were able to achieve mode-locking at various combinations of central wavelength and spectral width in our physics experiments (\Figref{Fig:06}, \Tabref{Tabl_2}), which confirms the validity of our preliminary calculations (\Figref{Fig:05}).

Exact real combinations of $\Delta\lambda$ and $\lambda_0$ realized in practice depend on the alignment of the optical components in the laser resonator, particularly on the Brewster angle settings and the aperture tuning in the optical setup.

\begin{table}
   \centering
   \caption{Experimental radiation spectral data in the mode-locking regime}
   \begin{tabular}{ccc}
   \hline
   \hline
       \textbf{The excessive} & \textbf{Central} & \textbf{Spectral width} \\
       \textbf{prism insertion} & \textbf{wavelength} & \textbf{of radiation $\Delta\lambda$} \\
       \textbf{depth $A$} & \textbf{of radiation $\lambda_0$} & \textbf{(at half-maximum)} \\
   \hline
   \hline
          3.07 mm & 808 nm & 37.5 nm \\
   \hline
          3.16 mm & 830 nm & 29.0 nm \\
   \hline
          3.38 mm & 850 nm & 27.5 nm \\
   \hline
          3.60 mm & 835 nm & 56.5 nm \\
   \hline
   \hline
   \end{tabular}
   \label{Tabl_2}
\end{table}

\section{CALCULATIONS OF GDD USING GEOMETRIC MODELING}
The accuracy of the calculated GDD parameters, obtained using the presented physical model, was further verified by means of geometrical modeling.

\begin{figure}[!bh]
  \centering
  \includegraphics[width=0.96\textwidth]{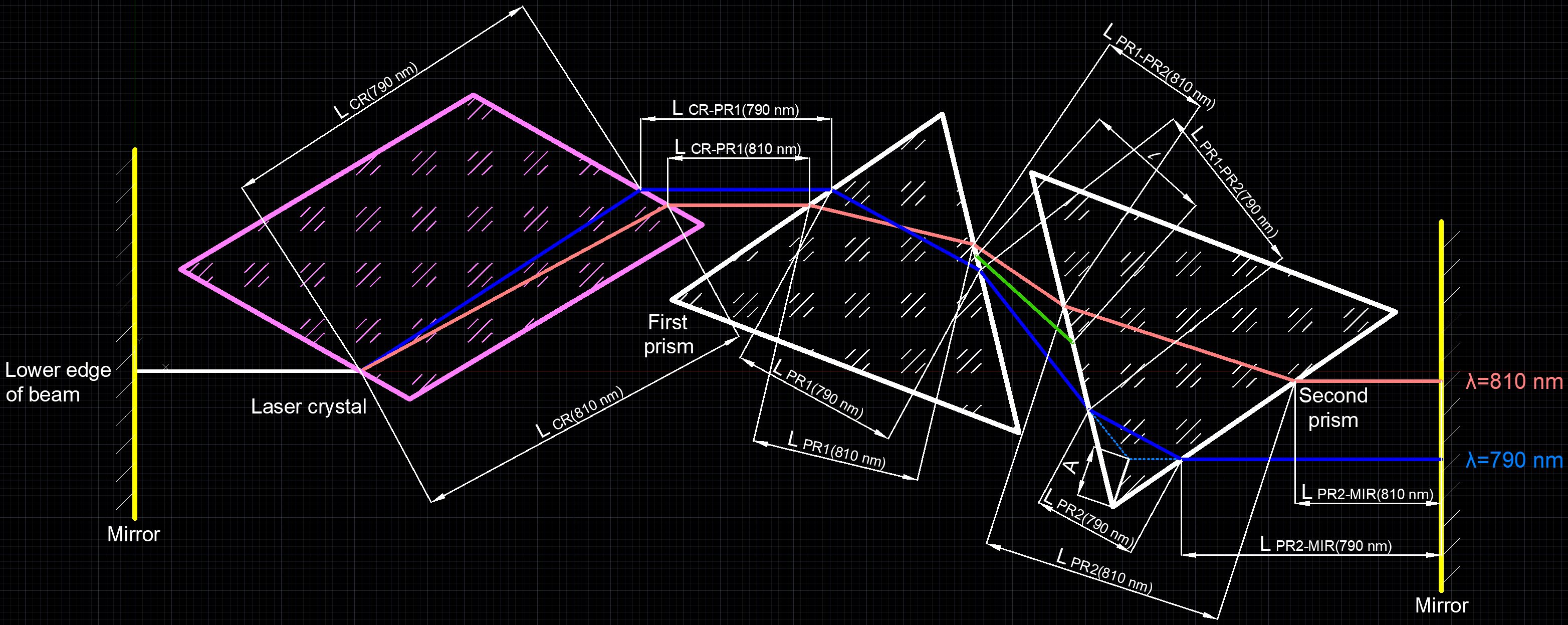}
  \caption{Geometric modeling of GDD compensation using a two-prism system with a distance between prisms $l$. Dashed lines indicate ray path prior to additional insertion of the second prism by depth $A$.}
  \label{Fig:07}
\end{figure}

As an example, the following parameters were considered: central wavelength of radiation $\lambda_0$ = 800 nm, spectral width $\Delta\lambda$ = 20 nm, the width of the crystal $t$ = 4.5 mm, the diameter of the laser beam $d$ = 2 mm, and the excessive prism insertion depth $A$ = 2.8 mm. For the lower edge of the laser beam, the trajectories of the extreme spectral components with wavelengths of 790 nm and 810 nm were constructed (\Figref{Fig:07}). The refraction angles at the interfaces between different media were calculated using Snell's law for each wavelength.

\onecolumngrid

\twocolumngrid

To fully compensate GDD the optical paths $P$ of all spectral components within the laser resonator should be equal to each other \cite{Marcuse1980}:

\begin{equation}\label{eq:20}
{P_1} = {P_2} = ... = {P_i},
\end{equation}

\begin{equation}\label{eq:21}
\begin{array}{l}
{P_i} = {L_{CR({\lambda _i})}} \cdot \left( {{n_{CR}}({\lambda _i}) - {\lambda _i} \cdot \frac{{d{n_{CR}}({\lambda _i})}}{{d{\lambda _i}}}} \right) + \\ + {L_{CR - PR1({\lambda _i})}} + \\
 + {L_{PR1({\lambda _i})}} \cdot \left( {{n_{PR}}({\lambda _i}) - {\lambda _i} \cdot \frac{{d{n_{PR}}({\lambda _i})}}{{d{\lambda _i}}}} \right) + \\ + {L_{PR1 - PR2({\lambda _i})}} + \\
 + {L_{PR2({\lambda _i})}} \cdot \left( {{n_{PR}}({\lambda _i}) - {\lambda _i} \cdot \frac{{d{n_{PR}}({\lambda _i})}}{{d{\lambda _i}}}} \right) + \\ + {L_{PR2 - MIR({\lambda _i})}},
\end{array}
\end{equation}

where $L_{(\lambda_i)}$ is the geometric path length of a spectral component with wavelength $\lambda_i$ on each straight section (see \Figref{Fig:07}): between optical elements ($L_{CR-PR1(\lambda_i)}$, $L_{PR1-PR2(\lambda_i)}$, $L_{PR2-MIR(\lambda_i)}$), within a laser crystal ($L_{CR(\lambda_i)}$) and within prisms ($L_{PR1(\lambda_i)}$, $L_{PR2(\lambda_i)}$). Refractive indices $n(\lambda_i)$ of crystal ($n_{CR}(\lambda_i)$) and prisms ($n_{PR}(\lambda_i)$) were determined using Eq. (\ref{eq:01}). The values of $\frac{dn(\lambda_i)}{d\lambda_i}$ were calculated for crystal and prism by the formula:

\begin{equation}\label{eq:22}
\frac{{dn({\lambda _i})}}{{d{\lambda _i}}} =  - \frac{{{\lambda _i}}}{{n({\lambda _i})}} \cdot \sum\limits_{j = 1}^3 {\frac{{\lambda _j^2 \cdot {B_j}}}{{{{\left( {\lambda _i^2 - \lambda _j^2} \right)}^2}}}} ,
\end{equation}

where the coefficients $B_j$ and $\lambda_j$ correspond to the material parameters listed in \Tabref{Tabl_1}.

The optical path difference between the spectral components at 790 nm and 810 nm within the resonator equaled zero for a distance between prisms $l$ = 608 mm, whereas the value calculated analytically using Eq. (\ref{eq:18}) was $l$ = 615 mm. Thus, the discrepancy between the mathematically calculated distance between prisms and the value obtained through geometrical modeling amounted to only 1.1 \%, which confirms the validity of the proposed approach.

\section{CONCLUSIONS}
In summary, this study presents a comprehensive theoretical and experimental analysis of how the configuration of the prism compressor affects the output parameters of a femtosecond laser system. The developed mathematical model enables precise calculation of the conditions required for mode-locking at specific central wavelengths and spectral widths. The results of geometrical modeling confirmed the accuracy of the analytical calculations, with a discrepancy of less than 1.1 \%. These findings offer a practical basis for optimizing dispersion compensation parameters to ensure stable generation of ultrashort pulses, which are essential for the implementation of dielectric laser acceleration (DLA) experiments.

\begin{acknowledgments}
The study is supported by the National Research Foundation of Ukraine under the program "Excellent Science in Ukraine" (project \# 2023.03/0182).
\end{acknowledgments}

\end{document}